\documentclass[reprint,aps,prb,amsmath,amssymb,dvipdfmx,twocolumn]{revtex4-2}
\usepackage{graphicx,amsmath,amssymb,bm,braket,color,comment,hyperref}
\hypersetup{colorlinks=true,citecolor=blue,linkcolor=blue,urlcolor=blue}
\begin{document}


\title{Triple-$q$ quadrupole-octupole order scenario for PrV$_2$Al$_{20}$}

\author{Takayuki Ishitobi}
\author{Kazumasa Hattori}%
\affiliation{%
 Department of Physics, Tokyo Metropolitan University,\\ 1-1, Minami-osawa, Hachioji, Tokyo 192-0397, Japan
}%

\date{\today}

\begin{abstract}
We propose novel triple-${\bm q}$ multipole orders as possible candidates for the two distinct low-temperature symmetry broken phases in quadrupolar system PrV$_2$Al$_{20}$.  An analysis of the experiment under [111] magnetic fields indicates that the {\it ferro} octupole moments in the lower temperature phase arise from the {\it antiferro} octupole interactions.  We demonstrate that the triple-${\bm q}$ multipole orders can solve this seemingly inconsistent issue. Anisotropies of quadrupole moments stabilize a triple-${\bm q}$ order, which further leads to the second transition to a coexisting phase with triple-${\bm q}$ octupole moments. The cubic invariant of quadrupole moments formed by the triple-${\bm q}$ components and the characteristic couplings with the octupole moments in their free energy play important roles. We analyze a multipolar exchange model by mean-field approximation and discuss the temperature and magnetic field phase diagrams. Many of the microscopic results, such as the number of phases and the magnitudes of critical fields in the phase diagrams, are qualitatively consistent with the experiments in PrV$_2$Al$_{20}$.
\end{abstract}

\pacs{Valid PACS appear here}
\maketitle


 { \it Introduction---}. Multipole degrees of freedom in spin-orbital coupled electron systems \cite{ReviewOrb1} possess a lot of intriguing physical phenomena such as multipolar ordering \cite{ReviewOrb2}, spin-orbital correlations \cite{Kugel,Jackeli}, Kondo effects \cite{Cox}, and superconductivity \cite{Chubkov,Nomoto}. An important feature of multipole physics is their anisotropy, and anisotropic properties such as magnetoelectric effects in the ordered phases have been actively studied in recent years \cite{Katsura,Hayami,Watanabe}. Among various multipoles, electric multipoles sometimes exhibit nontrivial mechanisms of their ordering and fascinating phenomena, e.g., in chiral nematic liquid crystals \cite{Fukuda}, Mott transitions \cite{Misawa,Zacharias}, and orbital transitions \cite{Hattori10}. Electric multipoles allow odd-order couplings in their free energy \cite{Cowley}, which can lead to novel phenomena forbidden by the time-reversal symmetry in magnetic systems. 
 
 Recently, Pr-based cubic compounds have been recognized to possess a possibility that non-Kramers $\Gamma_3$-doublet states are stabilized under the cubic $O_{\rm h}$ or $T_{\rm d}$ crystalline electric field (CEF) \cite{Yatskar,Tanida,OnimaruPrPb3,Kusanose19}. The $\Gamma_3$ doublet has no magnetic dipole moments while it has $\Gamma_3$ electric quadrupole and $\Gamma_2$ magnetic octupole moments. In the series of Pr-based 1--2--20 compounds, Pr$Tr_2X_{20}$ ($Tr$=Ti, V, Ir, $X$= Al, Zn, etc.), varieties of phenomena arising from the multipolar degrees of freedom have been observed experimentally \cite{reviewOnimaru,Matsubayashi,Sakai12,T-Sato,Taniguchi,Kittaka,Onimaru}. 
  In particular, PrV$_2$Al$_{20}$ shows many intriguing features such as non-Fermi liquid behavior \cite{Sakai11}, superconductivity \cite{Tsujimoto}, and multiphase diagram \cite{Shimura13,Shimura15,Shimura19}. Despite a number of theoretical works \cite{Hattori14,Tsuruta,Hattori16,Ishitobi,SBLee,KuboJPSJ,Freyer18,Freyer20,Inui}, most of them remain unsolved. 
  
 PrV$_2$Al$_{20}$ exhibits two consecutive (seemingly second-order) phase transitions at temperatures $T$$=$$T_1$$\sim$$0.8$ K and $T_2$$\sim$$0.7$ K \cite{Araki, Tsujimoto}.  The elastic anomalies suggest that the high-$T$ phase for $T_2$$<$$T$$<$$T_1$ is an antiferro $\Gamma_3$ quadrupolar order as familiar to other 1-2-20 compounds \cite{reviewOnimaru,Taniguchi,Kittaka,T-Sato,Onimaru}.  
  So far, a few studies have explored the nature of the low-$T$ phase.  Freyer {\it et al.} have proposed a possibility of $\Gamma_2$ octupole orders as the low-$T$ phase below $T_2$ by introducing biquadratic interactions between the quadrupole and the octupole moments \cite{Freyer18}.  
  Importantly, a recent unpublished report shows a hysteresis in the magnetostriction via the magnetic field ${\bm h}\parallel [111]$  sweep for $T$$<$$T_2$ \cite{Sakai-unpub,Patri19}.  This suggests the presence of the ferro $\Gamma_2$ octupole moment for $T$$<$$T_2$ \cite{Patri19}. As will be discussed below, the consistency between the simple ferro octupole scenario and the experiments is not evident when one considers the phase diagram under the magnetic fields.   
For many $f$-electron systems, clarifying the field-induced phases has played a crucial role in identifying the order parameters  \cite{Ho03,Joynt02,Cameron2016}.  In PrV$_2$Al$_{20}$, the phase diagram under the magnetic field \cite{Shimura13,Shimura15,Shimura19,Tsujimoto} is not compatible with that known for the N\'{e}el orders \cite{Hattori14}, exhibiting many unidentified phases.  Thus, theories must explain these field-induced phases and the double transitions in a unified way. 
 
Let us first demonstrate that the {\it closeness of $T_1$ and $T_2$ leads to a constraint on the critical field} $h^{111}_c$ for ${\bm h}\parallel$ [111], which rules out the simple ferro octupole order for $T<T_2$.
Suppose a second-order transition to a $\Gamma_3$ quadrupole ($\eta$) order occurs at $T$$=$$T_1$ and subsequently a $\Gamma_2$ ferro octupole ($\zeta$) order takes place at $T_2<T_1$. One can estimate the ground-state energy as $E\simeq -(a_1\eta^2+a_2\zeta^2)/2-\beta \tilde{h}_o\zeta$ for ${\bm h}=h^{111}(1,1,1)/\sqrt{3}$, where $a_{1,2}\sim T_{1,2}$, $\tilde{h}_o=(h^{111}/\sqrt{3})^3$ is the effective Zeeman field for the $\Gamma_2$ octupole moment; see Eqs.~(\ref{eq:Hamiltonian_Z}) and (\ref{eq:Zeeman}). Just below  $h_c^{111}$, $\zeta=$$\sqrt{1-\eta^2}$$\simeq$$1$$-$$\eta^2/2$ from the local constraint, and thus we obtain 
\begin{eqnarray}
E\simeq- \Bigl(\frac{a_2}{2}+\beta \tilde{h}_o\Bigr)+\frac{1}{2}(\beta \tilde{h}_o-a_1+a_2)\eta^2+\cdots 
  	\label{eq:formula}
.\end{eqnarray}
When the quadrupole order occurs at $h^{111}=11$ T \cite{Tsujimoto,Shimura19}, the coefficient of $\eta^2$ is negative: $a_1-a_2>\beta \tilde{h}_o\simeq 1.0$ K. However, this contradicts $a_1-a_2\sim T_1-T_2\sim 0.1$ K. 
This simple argument clearly shows that the simple ferro octupole order contradicts the double transitions in PrV$_2$Al$_{20}$. 

When the octupole interactions are antiferroic, the value of $h_c^{111}$ can be roughly estimated by Eq.~(\ref{eq:formula}) with $a_2\to -a_2$, which gives $h_c^{111}\sim 12.5$ T ($\beta \tilde{h}_o^{111}\sim 1.5$ K). This is very close to the experimental value.  It is not accidental but is the key to solving the problem.  Of course, usual antiferro octupole orders have no ferro octupole moment. 
However, this is not the case for the {\it triple}-${\bm q}$ quadrupole and octupole orders. We will show that the {\it triple}-${\bm q}$ quadrupole and octupole orders naturally explain the ferro octupole moment, in addition to the global phase diagrams. 
 The ideas are (i) the primary ordering wavevectors are those at the X point ${\bm q}$$=$${\bm q}_1$$\equiv$$(2\pi,0,0)$, ${\bm q}_2$$\equiv$$(0,2\pi,0)$, and ${\bm q}_3$$\equiv$$(0,0,2\pi)$ as expected in several Fermi surface studies \cite{Swatek,Iizuka,Nagasawa} and (ii) the order parameter for the high-$T$ (low-$T$) phase is a quadrupole (octupole) order. On the basis of the two assumptions, semiquantitatively consistent results with the experiments follow by the symmetry and the orbital character in PrV$_2$Al$_{20}$.

{\it Symmetry-allowed couplings---}.
First we discuss the free energy in this system from the point of view of symmetry. As noted above, the X point order parameters owing to the Fermi surface nesting \cite{Swatek,Iizuka,Nagasawa} are the key in our theory. Let us examine the free energy for the X point $\Gamma_3$ quadrupole orders with the wavevector ${\bm q}_l(l=1,2,3)$. Since the diamond structure formed by Pr$^{3+}$ ions can be regarded as two fcc lattices with one shifted by $(\frac{1}{4},\frac{1}{4},\frac{1}{4})$ to the other, the order parameters at the X points correspond to those for the decoupled fcc lattice model, where triple-${\bm q}$ quadrupole orders are known to be realized \cite{Tsune}. 
 Thus, one can extend the triple-${\bm q}$ theory in the fcc lattice to the present case.  
 The main source of this triple-$\bm{q}$ is the local cubic potential.  Denoting the $\Gamma_3$ quadrupole moment at the position ${\bm r}$ as ${\bm \eta}(\bm r)\equiv\eta(\bm r)[\cos \theta(\bm r), \sin \theta(\bm r)]$,
\begin{eqnarray}
	\mathcal F_{3}=-c_3\int d{\bm r} \eta^3({\bm r})\cos[3\theta(\bm{r})], \quad \label{eq:F3loc}
\end{eqnarray}
with $c_3>0$ \cite{param}. In terms of the Fourier components at  ${\bm q}$$=$${\bm q}_{l}$: ${\bm \eta}^{\lambda_l}_l\equiv \eta^{\lambda_l}_l(\cos \theta^{\lambda_l}_l, \sin \theta^{\lambda_l}_l)$
 with $\eta_l^{\lambda_l} \geq 0$ and the sublattice parity $\lambda_l$$=$$\pm$, the relevant terms for the triple-$\bm q$ orders are 
\begin{eqnarray}
	{\mathcal F}_{3}^{\rm X}=-12c_3\sum_{\lambda_1\lambda_2\lambda_3}\!\!\!\eta_1^{\lambda_1}\eta_2^{\lambda_2}\eta_3^{\lambda_3}\cos(\theta_1^{\lambda_1}+\theta_2^{\lambda_2}+\theta_3^{\lambda_3}).\label{eq:F3q}
\end{eqnarray}
 Note that ${\bm q}_1$$+$${\bm q}_2$$+$${\bm q}_3$$=$${\bm G}$  with ${\bm G}$ being the reciprocal lattice vectors and the two modes with $\lambda_\ell=\pm$ are degenerate at the X points. It is clear that ${\mathcal F}_{3}^{\rm X}<0$, when triple-${\bm q}$ orders with $\sum_{\ell=1}^3\theta_\ell^{\lambda_\ell}=0$ realize and the first-order transition temperature is higher than that for the single-${\bm q}$. Owing to the sublattice (inversion) symmetry, the combinations of $\lambda_{1,2,3}$ are $+++$ or $--+$ in Eq.~(\ref{eq:F3q}). Below we focus on the $+++$ combination, since the others represent just different domains of the $+++$ for most of the phases realized in the microscopic calculations shown later.

Now we discuss one of the main aspects: The double transitions in PrV$_2$Al$_{20}$, 
on the basis of the symmetry-allowed couplings.
Since the triple-$\bm q$ state is a {\it partial-order} state including disordered sites \cite{Tsune}, it is natural to expect there is a further symmetry breaking at the lower temperature.  Indeed, a triple-${\bm q}$ octupole order with {\it finite ferro octupole moments} naturally emerges when just considering {\it local repulsion} between the quadrupole and the octupole moments:
\begin{eqnarray}
	\mathcal F_{4}=\int d{\bm r}  \left[c_4 \eta^4({\bm r})+c'_4 \zeta^4({\bm r})+c''_4\eta^2({\bm r}) \zeta^2({\bm r}) \right], \quad \label{eq:F4}
\end{eqnarray}
where $c_4$'s are constants\cite{param} and $\zeta({\bm r})$ is the $\Gamma_2$ octupole moment. The last term in Eq.~(\ref{eq:F4}) represents the repulsion between 
${\bm \eta}({\bm r})$ and $\zeta({\bm r})$.
 This, in its Fourier form, includes two key couplings between the parity-even quadrupole $\eta^+_l$ and the octupole $\zeta^+_l$ at the X points and among $\eta^+_l$, $\zeta^+_l$, and the uniform octupole moment $\zeta_0^+$.
The first is the main source of the double transitions and given by 
\begin{eqnarray}
{\mathcal F}_{4}^{(1)}=8c''_4\eta_1^{+}\eta_2^{+}\zeta_1^{+}\zeta_2^{+} \cos(\theta_1^{+}-\theta_2^{+})+ (123: cyclic),\ \   \label{eq:third2-1}
\end{eqnarray}
which favors triple-${\bm q}$ octupole orders under the triple-${\bm q}$ quadrupole states. This means that if the transition at $T=T_1$ is the X point quadrupole order, it is reasonable that the triple-${\bm q}$ octupole order takes place below $T_2$ with the help of Eq.~(\ref{eq:third2-1}). Second, the coupling between the $\Gamma$ and the X points exists:
\begin{eqnarray}
{\mathcal F}_{4}^{(2)}=8c''_4 \eta_1^{+}\eta_2^{+}\zeta_3^{+}\zeta_0^{+} \cos(\theta_1^{+}-\theta_2^{+}) + (123: cyclic).\ \ \ \ 
\label{eq:third2-2}
\end{eqnarray}
 This term is one of the main results in this work and explains  {\it finite ferro octupole moments under triple-${\bm q}$ quadrupole-octupole orders.} Note that again the condition ${\bm q}_1$$+$${\bm q}_2$$+$${\bm q}_3$$=$${\bm G}$ plays an important role.

{\it Model---}. So far, we have demonstrated that the triple-$\bm q$ orders can qualitatively explain the double transitions and the finite ferro octupole moments in PrV$_2$Al$_{20}$. Now we construct an exchange model for PrV$_2$Al$_{20}$. 
Indeed, the estimated entropy at $2$ K is $\sim R\ln 2$ in the specific heat studies \cite{Sakai11,Tsujimoto}.  Thus a localized model is a reasonable starting point to study the ordered phases. 
Pr$^{3+}$ ions with 4f$^2$ configuration form a diamond structure, and the local point group is $T_{\rm d}$ \cite{reviewOnimaru}. The nine states in the total angular momentum $J$$=$$4$ are split by the CEF into $\Gamma_3$(0 K), $\Gamma_5$($\epsilon_5$=40 K), $\Gamma_4$($\epsilon_4$=77 K) triplets, and $\Gamma_1$($\epsilon_1$=184 K) singlet \cite{[{The CEF levels $\epsilon_{1-5}$ are calculated by using the CEF parameters obtained in the magnetization experiments in }] Araki}. In order to take into account the local cubic anisotropy Eq.~(\ref{eq:F3loc}), the minimal model must include the nonmagnetic $\Gamma_1$ singlet. The other excited states do not enter the low-energy physics and are safely ignored.

We denote $\Gamma_3(u,v)$ and $\Gamma_1(s)$ states at the site $i$ in the local basis $(|u\rangle_i,|v\rangle_i,|s\rangle_i)$. The $\Gamma_3$ quadrupole ${\bm \tau}_i$$=$$(\tau_i^u, \tau_i^v)$ and the $\Gamma_2$ octupole $t_i$ operators are expressed as 
\begin{equation}
\tau^u_i:\left(
\begin{aligned}
&\ \sigma_z &\!\!\!\!\begin{split}
	a\\[-1mm]
	0 
\end{split}\\[-1mm]
&a\ \ 0 &\! 0
\end{aligned}
\right), \ 
\tau^v_i:\left(
\begin{aligned}
&\!\!-\!\sigma_x &\!\!\!\!\begin{split}
	0\\[-1mm]
	a 
\end{split}\\[-1mm]
&0\ \ a &\! 0
\end{aligned}
\right), \
t_i:\left(
\begin{aligned}
&\ \sigma_y &\!\!\!\!\begin{split}
	0\\[-1mm]
	0 
\end{split}\\[-1mm]
&0\ \ 0 &\! 0
\end{aligned}
\right), 	
\label{eq:multipoles2}
\end{equation}
where $\sigma_{x,y,z}$ is the Pauli matrix and $a$$=$$\sqrt{35}/2$ \cite{Hattori14}. 
In addition to the inclusion of the $\Gamma_1$ state, anisotropic interactions are also important \cite{Kugel,Kubo17}. We take into account all the possible exchange interactions for the nearest-neighbor (NN) and 
the next-NN (NNN) pairs. The Hamiltonian is $\mathcal H=\epsilon_1 \sum_i|s\rangle_i {}_i\langle s|+\mathcal H_{\rm int}+\mathcal H_{\rm Z}$, where
\begin{align}
	{\mathcal H}_{\rm{int}} &= \!\!\sum_{i,j}[J_{ij}{\bm \tau_i}\cdot {\bm \tau}_j + J_{ij}^{o} t_i t_j+K_{ij}
	(\tilde{\tau}^u_{ij} \tilde{\tau}^u_{ji}-\tilde{\tau}^v_{ij} \tilde{\tau}^v_{ji})],\label{eq:Hamiltonian_int}\\
	 {\mathcal H}_{\rm{Z}} &= -\sum_i[\alpha \tilde{\bm h} \cdot {\bm \tau}_i + \beta \tilde{h}_{o}t_i].\label{eq:Hamiltonian_Z}
\end{align}
Here, the $i,j$ sum in Eq.~(\ref{eq:Hamiltonian_int}) is restricted up to
the NNN sites. We will use the notations $J_{1(2)}$, $J_{1(2)}^{o}$, and $K_{1(2)}$ for NN(NNN) interactions in $J_{ij}$, $J^{o}_{ij}$, and $K_{ij}$, respectively ($K_1=0$ by the symmetry). The last term in Eq.~(\ref{eq:Hamiltonian_int}) is the anisotropic one: $\tilde{\tau}^{u,v}_{ij}$$\equiv$${\bm n}^{u,v}_{ij} $$\cdot$${\bm \tau}_i$ and ${\bm n}^u_{ij}$$=$$(\sin\phi_n$, $-\cos\phi_n)$ and 
${\bm n}^v_{ij}$$=$$(\cos\phi_n$, $\sin\phi_n)$ with $\phi_n$$=$$\frac{2n\pi}{3}$, $n$$=$$0$, $1$, and $2$, when the bond $i$-$j$ lies on the $xy$, $yz$, and $zx$ plane, respectively \cite{Kubo17}. Under magnetic fields ${\bm h}$$=$$(h_x,h_y,h_z)$, the quadrupole and octupole moments couple to ${\bm h}$ via the pseudo Zeeman coupling Eq.~(\ref{eq:Hamiltonian_Z}); $\tilde{\bm{h}}$ and $\tilde{h}_o$ are given by 
\begin{eqnarray}
	\tilde{\bm{h}}=(2h_z^2-h_x^2-h_y^2, \sqrt{3}(h_x^2-h_y^2)),\quad  \tilde{h}_{o}=h_xh_yh_z,\ \ \ \label{eq:Zeeman}
\end{eqnarray}
 and $\alpha=(g\mu_{\rm B})^2$$[7/(3\epsilon_4)$$-$$1/\epsilon_5]\simeq$ 0.0015 KT$^{-2}$ \cite{Hattori14}, and $\beta=(g\mu_{\rm B})^3$$\sqrt{3}$$[7/\epsilon_4^2$$+$$15/\epsilon_5^2$$+$$14/(\epsilon_4\epsilon_5)]\simeq$ 0.004 KT$^{-3}$. Here, $\mu_{\rm B}$ is the Bohr magneton and $g$$=$$4/5$.
 
 Among the wide range of parameter space spanned by $J_{1,2}$, $J_{1,2}^o$, and $K_2$, we find that the parameter set $|J_2|,|J_2^{o}|$$>$$|J_1|,|J_1^{o}|,K_2$ with $J_1, J_1^{o}<0$ and $K_2>0$ are well compatible with the experimental results in PrV$_2$Al$_{20}$.
 Indeed this choice is consistent with the X point nesting \cite{Swatek,Iizuka,Nagasawa} and reflects the realistic situation. 
The results for $J_1>0$ are also comparable with the experimental data, but here we restrict ourselves to the $J_1<0$ case, and the full analysis will be published elsewhere \cite{IshitobiFullPaper}. Further experimental information is needed to fix the sign, although this might suggest that $J_1$, which favors the N\'{e}el state, is not important for PrV$_2$Al$_{20}$.   

{\it Stability against incommensurate orders---}.
Since the assumption (i) about the ordering wavevectors plays a crucial role in our theory, let us discuss this point first. Analyzing the Fourier transform of the exchange coupling $J_{\bm q}$ derived from Eq.~(\ref{eq:Hamiltonian_int}), one notices that the ordering vector ${\bm q}_1^*$ is incommensurate: 
$\bm{q}_1^* \simeq {\bm q}_1-(\delta q,0,0)$ with $\delta q=|J_1|/(2 J_2+K_2)$ for small $|J_1|$, at which 
 $J_{\bm{q}_1^*}\simeq -2J_2-4K_2-\delta q |J_1|/4$. In order to realize the X point triple-${\bm q}$ quadrupole order at $T=T_1$ against the incommensurate spiral states, the third-order free energy  gain [Eq.~(\ref{eq:F3loc})] must overwhelm the energy cost $\delta q |J_1|/4$. This is possible because the eigenmode of $J_{{\bm q}_\ell}$ is that for $\theta_{\ell}^+=2\pi\ell/3$ \cite{Tsune}, leading to the maximum gain of Eq.~(\ref{eq:F3loc}).
 The condition is estimated by the standard Landau theory for the triple-${\bm q}$ transition at $T=T_1$ and 
\begin{eqnarray}
	\frac{T_1}{2}-2J_2-4K_2-\frac{\delta q |J_1|}{4}=\frac{c_3^2}{2c_4}-\frac{J_1^2}{4(2J_2+K_2)}>0,\ \ \ \ 
\end{eqnarray}
where $c_3 \simeq 0.021$ and $c_4\simeq 0.095$ K in PrV$_2$Al$_{20}$ for $T=1$ K \cite{param}. This leads to 
$J_1^2/(2J_2+K_2)<0.0090$ K. For the parameters giving $T_{1}\sim 0.8$ K, e.g., $J_2=0.16$ and $K_2=0.02$ K, the condition becomes $|J_1|<\sqrt{0.0090 \cdot 0.34}	\sim 0.055 \ {\rm K}\sim J_2/3$ (the condition is relaxed if the third-neighbor couplings are taken account).
Thus the anisotropy [Eq.~(\ref{eq:F3loc})] {\it locks} the ordering vector at ${\bm q}_\ell$ for small $|J_1|$. As long as the leading single-${\bm q}$ instability occurs near ${\bm q}_{\ell}$, the essential results are invariant. 

\begin{figure}[t!]
\begin{center}
\includegraphics[width=0.5\textwidth]{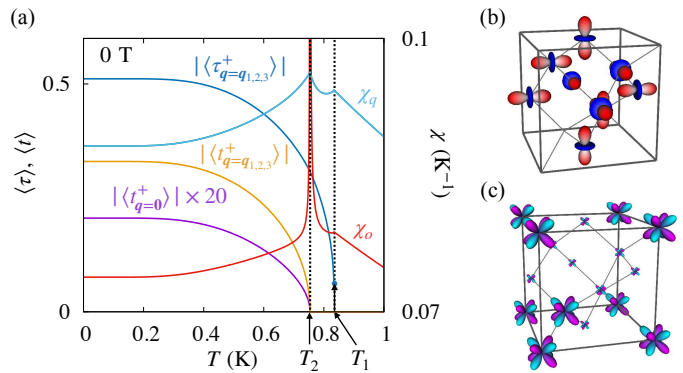}
\end{center}
\vspace{-15pt}
\caption{
(Color online). (a) $T$ dependence of uniform quadrupole ($\chi_q$) and octupole ($\chi_o$) susceptibility for $|{\bm h}|=0$. 
Absolute value of quadrupole $\langle \bm{\tau}^+_{\bm q}\rangle$ and octupole $\langle t^+_{\bm q}\rangle$ moments with + parity at the X points and uniform octupole moment $\langle t^+_{\bm q= {\bm 0}}\rangle$ are shown. There is a small jump in $\langle \bm{\tau}^+_{\bm q}\rangle$ at $T=T_1$. (b) Quadrupoles pattern for $T<T_1$ and (c) octupoles for $T<T_2$ \cite{[{Illustrations of order-parameter configurations have been made with the Mayavi software package. }]  mayavi}. 
}
\label{fig:chiQ}
\end{figure}

{\it Results---}.
The mean-field calculations are carried out by assuming eight-site orders in the cubic unit cell, which corresponds to the ordering vectors at the X points. Since the (putative) transition temperature $T_{\rm Q}$ for the single-${\bm q}$ at the X points is proportional to $J_2$$+$$2|K_2|$, we set $J_2$$+$$2|K_2|$$=$$0.2$ K, which gives $T_{\rm Q}\sim 0.8$ K$\sim T_1$. 


\begin{figure*}[t]
\begin{center}
\includegraphics[width=0.95\textwidth]{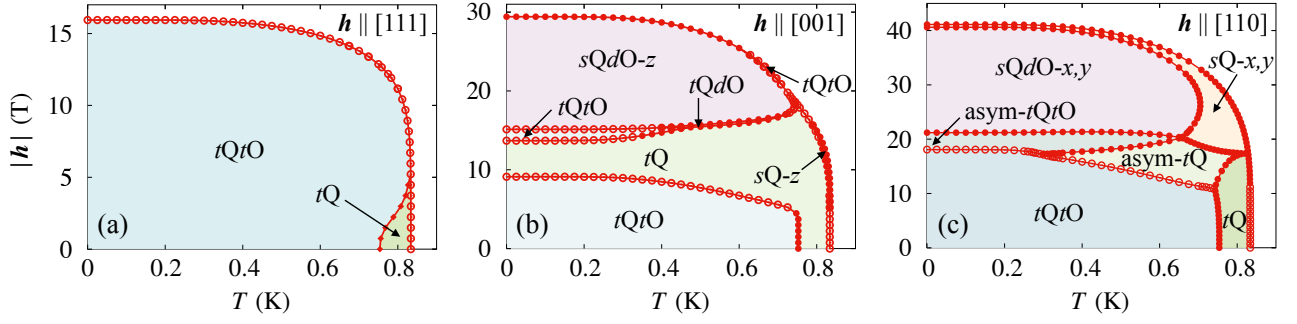}
\end{center}
\vspace{-15pt}
\caption{
(Color online). $T$--$\bm{h}$ phase diagrams for $J_1=J_1^o=-0.02$, $J_2=0.154$, $K_2=0.023$, and $J_2^o=0.195$ K. (a) ${\bm h}$ $\parallel$ $ [111]$, (b) $[001]$, and (c) $[110]$. The filled (open) circles $\bullet$ ($\circ$) represent  (dis)continuous transitions, and the filled diamonds $\blacklozenge$ in (a) indicate a cross over.  }
\label{fig:PhaseDiagrams}
\end{figure*}

In Fig.~\ref{fig:chiQ}, $T$ dependence of uniform quadrupole susceptibility $\chi_q$ is shown 
for $|{\bm h}|= 0$, $J_1=J_1^o=-0.02$, $J_2=0.154$, $K_2=0.023$, and $J_2^o=0.195$ K. This parameter set indeed leads to the double transitions, where the high-temperature phase  for $T_2<T<T_1$ is the triple-$\bm{q}$ quadrupole ($t$Q) order with a finite expectation value $\langle \tau^+_{{\bm q}_{1,2,3}} \rangle$ through the weak first-order transition, while for $T<T_2$ it is the triple-${\bm q}$ quadrupole-octupole one ($t$Q$t$O) with $\langle \tau^+_{{\bm q}_{1,2,3}} \rangle, \langle t^+_{{\bm q}_{1,2,3}} \rangle \neq 0$. In $t$Q phase, the quadrupole moments vanish at two sites in the eight-site unit cell as shown in Fig.~\ref{fig:chiQ}(b). This reflects in the measurable $\chi_q$ via ultrasonic experiments; even for $T_2<T<T_1$ $\chi_q$ increases as lowering $T$, since the $\Gamma_3$ degeneracy remains at 1/4 of the sites.  For $T<T_2$, a  finite ferro octupole moment $\langle t_{{\bm q}={\bm 0}}\rangle$ emerges as a part of the order parameters as shown in Fig.~\ref{fig:chiQ}(c), which can be explained by Eq.~(\ref{eq:third2-2}) together with the diverging uniform octupole susceptibility $\chi_o$ at $T=T_2$ in Fig.~\ref{fig:chiQ}(a).

The ${\bm h}$--$T$ phase diagrams are shown in Fig.~\ref{fig:PhaseDiagrams} for $\bm{h} \parallel [111]$, [001], and [110]. 
For [111], $\tilde{\bm h}={\bm 0}$, $\tilde{h}_o=$$|{\bm h}|^3/(3\sqrt{3})$, and the uniform octupole moment is induced. Then, the second transition at $T_2$ smears out and becomes a crossover for $|\bm{h}|>0$ as shown in Fig.~\ref{fig:PhaseDiagrams}(a). Although the primary order parameters are antiferroic, hysteresis under the field sweep exists as observed in the experiment \cite{Patri19,Sakai-unpub}.


\begin{figure}[b!]
\begin{center}
\includegraphics[width=0.45\textwidth]{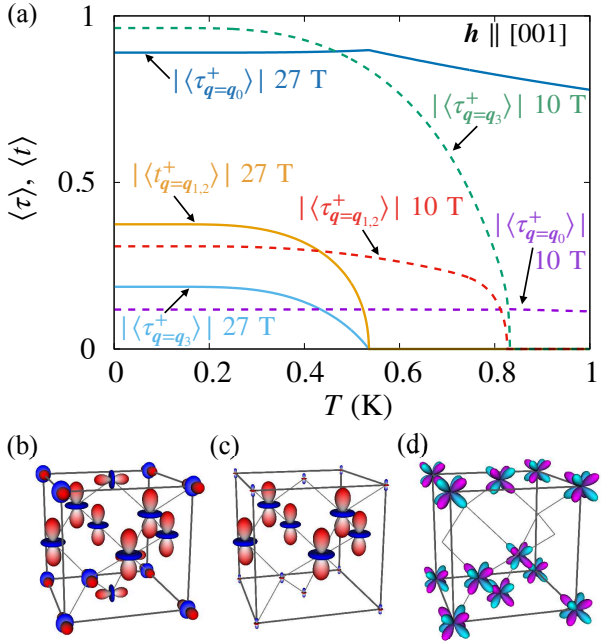}
\end{center}
\vspace{-15pt}
\caption{
(Color online). (a) $T$ dependence of the quadrupole $\langle \bm{\tau}^+_{{\bm q}}\rangle$ and octupole $\langle t^+_{\bm{q}}\rangle$ moments for $|{\bm h}|=27$ T (lines) and 10 T (dashed lines) along $\bm{h} \parallel [001]$. 
Ordered patterns are drawn for (b) quadrupoles in $t$Q under fields, (c) quadrupoles in $s$Q$d$O-$z$, and (d) octupoles in $s$Q$d$O-$z$ \cite{mayavi}.}
\label{fig:chiQ2}
\end{figure}

For ${\bm h}$ $\parallel$ $[001]$, $\tilde{\bm h}$$=$$2|{\bm h}|^2(1,0)$ and $\tilde{h}_o=0$. Then the uniform quadrupole moment $\eta_0^+$$(\cos\theta^+_0$, $\sin\theta^+_0)=(\eta^+_0,0)$ is induced and ``polarizes'' for large $|{\bm h}|$. The uniform moment couples with single-${\bm q}$ quadrupole moments via the local cubic potential Eq. (\ref{eq:F3loc}) as 
\begin{eqnarray}
	{\mathcal F}_{3}^{\rm X\Gamma}=-6c_3\sum_{\ell=1}^3\sum_{\lambda_{\ell}=\pm}\eta_\ell^{\lambda_\ell}\eta_\ell^{\lambda_\ell}\eta_0^{+}\cos(2\theta_\ell^{\lambda_\ell}+\theta_0^{+}).\label{eq:F3q2}
\end{eqnarray}
This favors the configurations in which $\eta_3^+$ with $\theta_3^+=0$ in $t$Q($t$Q$t$O) is larger than $\eta_{1,2}^+$; see Figs.~\ref{fig:chiQ2}(a) and \ref{fig:chiQ2}(b).
As a typical example, the dashed lines in Fig.~\ref{fig:chiQ2}(a) show $T$ dependence of $|\langle \tau^+_{{\bm q}_3}\rangle|> |\langle \tau^+_{{\bm q}_{1,2}}\rangle|$. A tiny interval between the onset of $\langle \tau^+_{{\bm q}_3}\rangle$ and $\langle \tau^+_{{\bm q}_{1,2}}\rangle$ corresponds to a single-${\bm q}$ state($s$Q-$z$). 
At high fields, a double-$\bm q$ octupole with 
 a single-${\bm q}$ quadrupole order for $\theta_3^+$$=$$0$ occurs as denoted by $s$Q$d$O-$z$, where ``$z$'' represents the direction of ${\bm q}$ for the quadrupole sector. The emergence of $s$Q$d$O-$z$ is understood by Eq.~(\ref{eq:F4}), which includes
\begin{eqnarray}
{\mathcal F}_{4}^{(3)}=8c''_4\eta_0^{+}\eta_1^{+}\zeta_2^{+}\zeta_3^{+} \cos(\theta_1^{+}-\theta_0^{+}) +(123: cyclic). \ \ \ \ 
\label{eq:third2-3}
\end{eqnarray}
As shown in Fig.~\ref{fig:chiQ2}(a), the double-${\bm q}$ octupole orders induce the single-${\bm q}$ quadrupole moments. Note that the latter linearly increases near the transition temperature in Fig.~\ref{fig:chiQ2}(a). In the $s$Q$d$O-$z$ phase, half of the quadrupole moments are enhanced, while the other half shrinks owing to ${\mathcal H}_{\rm Z}$ [Eq.~(\ref{eq:Hamiltonian_Z})]. Then the octupoles emerge at the sites where the quadrupoles are small; see Figs.~\ref{fig:chiQ2}(c) and \ref{fig:chiQ2}(d). 
In the intermediate fields, $t$Q state without octupole moments survives even at $T=0$. Between $t$Q and $s$Q$d$O-$z$, there is a tiny region of $t$Q$d$O phase at high temperature, while it is replaced by reentrant $t$Q$t$O state at low temperatures. We also note that there are very small regions where $s$Q-$z$ or $t$Q$t$O state appears at finite fields. This suggests that there are many competing orders having very close free energy with each other.

For ${\bm h}$ $\parallel$ $[110]$, $\tilde{\bm h}=$$-$$|{\bm h}|^2(1,0)$ and $\tilde{h}_o=0$.
Then $\theta_0^+=\pi$ uniform quadrupole moment is induced, which, in turn, favors the quadrupole orders with ${\bm q}_{1,2}$ rather than with ${\bm q}_3$ owing to Eq. (\ref{eq:F3q2}). In fact, $s$Q-$x,y$ and $s$Q$d$O-$x,y$ states appear in Fig.~\ref{fig:PhaseDiagrams}(c). For these states, the quadrupole order parameters rotate from their eigenmode direction $\theta_\ell^+=2\pi\ell/3$. In the asymmetric-$t$Q($t$O) state, the symmetry between $\bm{q}_1$ and $\bm{q}_2$ is broken and the magnitudes of the three triple-${\bm q}$ components differ. The appearance of such low-symmetry states is the consequence of frustration between the quadrupole moments for $\bm{q}=\bm{0}$ and $\bm{q}_\ell$  through Eqs.~(\ref{eq:F3q}), (\ref{eq:Hamiltonian_Z}), and (\ref{eq:F3q2}). This is unique for ${\bm h}\parallel$ [110] and is never realized for ${\bm h}\parallel$ [001] and [111].

{\it Discussions---}. We have demonstrated that the triple-${\bm q}$ quadrupole (and octupole) orders can be a promising candidate of the order parameter in PrV$_2$Al$_{20}$. Let us now discuss the difference between our theory and simple quadrupole/octupole order scenarios. 
 As noted earlier, starting from the antiferroic interactions and the set up $T_1-T_2\sim 0.1$ K, the critical field of $t$Q$t$O phase $h_c^{111}\sim 15$ T in Fig. \ref{fig:PhaseDiagrams} is consistent with the value in the experiments \cite{Tsujimoto,Shimura19}.  This is one of the important points in our theory in contrast to the simple ferro octupole order. 
For other field directions, the present results also agree with the data available at present overall: The number of the field-induced phases and the value of the critical field $\sim$ 27 T for ${\bm h}$ $\parallel$ [001] \cite{Shimura13,Nakanishi,Shimura15} and much higher for [110] \cite{Shimura19}. Moreover, the disappearance of $T_2$-anomaly at $\sim$ 7 T for [001] in the ultrasonic experiment \cite{Nakanishi} is consistent with the phase diagram in Fig.~\ref{fig:PhaseDiagrams}(b). In addition to this, the tiny phase between $t$Q and $s$Q$d$O-$z$ in our theory corresponds to the small phase between the high and intermediate phases with the first-order discontinuity at low temperatures in the magnetization experiment \cite{Shimura13}. 

For ${\bm h}$ $\parallel$ [110], there are several anomalies in magnetoresistance as briefly reported in Ref.~[\onlinecite{Shimura19}]. In this sense, the present result in Fig. \ref{fig:PhaseDiagrams}(c) is qualitatively consistent with the experiment. This seems to be another important point, since the phase diagram for ${\bm h}\parallel$ [110] within the NN model consists of only one antiferro quadrupole ordered state \cite{Hattori14,Hattori16}. 
For complete understanding, the experimental identification of the primary order parameter and the detail data for ${\bm h}\parallel$ [110] are highly desired. As the experimental checkpoints at $|\bm{h}|=0$, we note that there must be elastic softening of the $\Gamma_3$ mode in $t$Q state and the $C_3$ symmetry preserved in $t$Q$t$O state. The latter can be examined by transport measurements in addition to x-ray or neutron scattering experiments.

{\it Summary---}. We propose a multipolar triple-${\bm q}$ state as a promising order parameter for PrV$_2$Al$_{20}$, explaining the double transitions, critical fields, and phase diagrams under magnetic fields on the basis of a minimal quadrupole-octupole exchange model. The triple-${\bm q}$ states with partially disordered sites naturally explain the double transitions. And finite ferro octupole moments emerge in the triple-${\bm q}$ quadrupole and octupole ordered state. We believe our work stimulates further studies and opens a way to a deeper understanding of the multipole orders in various systems.

{\it Acknowledgement---.} The authors thank H. Tsunetsugu for fruitful discussions. This work was supported by a Grant-in-Aid for Scientific 
Research (Grants No.~16H04017, No. 18K03522, and No. 21H01031) from the Japan 
Society for the Promotion of Science.

\bibliography{Ref}

\begin{thebibliography}{55}
\bibitem [{par()}]{param}%
  \BibitemOpen
  \href@noop {} {}\bibinfo {note} {$c_3$ and $c_4$s can be obtained from the
  local CEF Hamiltonian and $c_3=\alpha T/2$, $c_4=T(1-\alpha)/12$,
  $c'_4=T/12$, and $c''_4=T(2-\alpha)/12$ with $\alpha=35T/(4\epsilon_1)\ll 1$
  for $T\simeq T_{1,2}\sim 1$ K in PrV$_2$Al$_{20}$.}\BibitemShut {Stop}%
\end{thebibliography}

\end{document}